\documentclass{article}
\usepackage[utf8]{inputenc}
\usepackage{bm}
\usepackage{amsmath}
\usepackage{amssymb}
\usepackage{graphicx}
\usepackage{titlesec}
\usepackage{comment}
\usepackage{subcaption}
\usepackage{arxiv}
\usepackage[title]{appendix}
\usepackage{multicol}

\newcommand{\cL}{{\cal L}}
\newcommand{\cX}{{\cal X}}
\newcommand{\vc}{ {\bf{c}}}
\newcommand{\vC}{ \bf{C}}
\newcommand{\vf}{ {\bf{f}}}
\newcommand{\vK}{ {\bf{K}}}
\newcommand{\vP}{ {\bf{P}}}

\newcommand{\cY}{{\cal Y}}
\newcommand{\cQ}{{\cal Q}}
\newcommand{\NN}{{\cal N}}
\newcommand{\bigO}{{\cal O}}
\renewcommand {\ij}{_{\c \d}} 
\newcommand{ \offset}{c\ij}
\newcommand{\vmu}{\boldsymbol{\mu}}
\newcommand{\vSigma}{\boldsymbol{\Sigma}}
\newcommand{\vy}{\boldsymbol{y}}
\newcommand{\vx}{\boldsymbol{x}}
\newcommand {\vX}{\boldsymbol{X}}
\newcommand{\vz}{\boldsymbol{z}}
\newcommand{\vtheta}{\boldsymbol{\theta}}
\newcommand{\vgamma}{\boldsymbol{\gamma}}
\newcommand{\vrho}{\boldsymbol{\rho}}
\newcommand {\n}{^{(n)}}
\newcommand{\gum}{GUM}
\newcommand{\RR}{\mathbb{R}}
\newcommand{\EE}{\mathbb{E}}

\newcommand{\nonlinear}{non-linear }

\renewcommand{\input}{regressor}
\newcommand{\inputs}{regressors}
\renewcommand{\output}{dependent variable}

\newcommand {\beq}{\begin{equation}}
\newcommand {\eeq}{\end{equation}}
\newcommand {\beqarray}{\beq  \left\{ \begin{array}{cll}}
\newcommand {\eeqarray}{ \end{array} \right.  \eeq}
\newcommand{\const}{\text{const}}
\newcommand {\gEy}{g(\mathbb E(y\n)))} 
\renewcommand{\c}{i}
\renewcommand{\d}{j}
\newcommand {\DM}{\bm \Phi} 
\newcommand{\estar}{^{\star}}
\newcommand{\estarT}{^{\star T}}
\newcommand{\dstar}{\d \estar_{\c}} 

\newcommand {\f}{f\ijl}

\newcommand {\ijstar}{ _{\c,\dstar} }
\newcommand {\ijl}{_{k(\c \d l)}}  

\newcommand {\muij}{\vmu\ij}
\newcommand {\Kcd}{\vK\ij}

\newcommand {\hp}{\vgamma}
\newcommand {\map}{^{\text{MAP}}}
\newcommand {\mustar}{\vmu\estar}
\newcommand {\mustartilde}{\bm {\tilde \mu}\estar}
\newcommand {\Kstar}{\vK\estar}
\newcommand {\Pstar}{\vP\estar}
\newcommand {\PstarT}{\vP\estarT}
\newcommand {\DMstar}{\DM\estar}

\newcommand {\DMstartilde}{\bm {\tilde \Phi}\estar}
\newcommand {\DMstartildeT}{\bm {\tilde \Phi}\estarT}
\newcommand {\Sigmatilde}{\tilde {\vSigma}}
\newcommand {\freepars}{\tilde \vtheta} 
\newcommand {\freestar}{\freepars\estar}
\newcommand {\freestarT}{\freepars^{\star T}}
\newcommand {\Udif}{(\freepars\estar {-}\mustartilde)}

\titleformat{\paragraph}
{\normalfont\normalsize\bfseries}{\theparagraph}{1em}{}
\titlespacing*{\paragraph}
{0pt}{3.25ex plus 1ex minus .2ex}{1.5ex plus .2ex}

\title{Non-linear regression models for behavioral and neural data analysis}
\author{Vincent Adam\\
PROWLER.io\\
Cambridge, UK\\
\texttt{vincent.adam@prowler.io} \\
\And Alexandre Hyafil \\
Centre de Recerca Matemàtica \\
Campus UAB Edifici C, 08193 Bellaterra, Spain \\
and Center for Brain and Cognition \\
Universitat Pompeu Fabra, 08018 Barcelona, Spain
}

\date{November 2019}

\begin{document}

\maketitle

\begin{abstract}
Regression models are popular tools in empirical sciences to infer the influence of a set of variables onto a dependent variable given an experimental dataset. In neuroscience and cognitive psychology, Generalized Linear Models (GLMs) -including linear regression, logistic regression, and Poisson GLM- is the regression model of choice to study the factors that drive participant's choices, reaction times and neural activations. These methods are however limited as they only capture linear contributions of each regressors. Here, we introduce an extension of GLMs called Generalized Unrestricted Models (GUMs), which allows to infer a much richer set of contributions of the regressors to the dependent variable, including possible interactions between the regressors. In a GUM, each regressor is passed through a linear or nonlinear function, and the contribution of the different resulting transformed regressors can be summed or multiplied to generate a predictor for the dependent variable. We propose a Bayesian treatment of these models in which we endow functions with Gaussian Process priors, and we present two methods to compute a posterior over the functions given a dataset: the Laplace method and a sparse variational approach, which scales better for large dataset. For each method, we assess the quality of the model estimation and we detail how the hyperparameters (defining for example the expected smoothness of the function) can be fitted. Finally, we illustrate the power of the method on a behavioral dataset where subjects reported the average perceived orientation of a series of gratings. The method allows to recover the mapping of the grating angle onto perceptual evidence for each subject, as well as the impact of the grating based on its position. Overall, GUMs provides a very rich and flexible framework to run nonlinear regression analysis in neuroscience, psychology, and beyond.
\end{abstract}

\keywords{regression \and Gaussian process\and neuroscience}

\section{Introduction}

\subsection{Regression models for data analysis}

Research questions in neuroscience and cognitive science often imply to empirically assess the factors that determine an observed neural activity or behavior in controlled experimental environments. Exploratory analyses on such datasets are typically performed using regression analyses -  where the measured data (e.g. neural spike count, subject choice, pupil dilation) is regressed against a series of factors (sensory stimuli, experimental conditions, history of neural spiking or subject choices, etc.). \\

A method of choice is the use of generalized linears models (GLMs \cite{mccullagh1989}) where the  \output{} is predicted from a linear combination of the factors. More formally, GLMs are regression models from the input space $\cX$ to the output space $\cY$ specifying a conditional distribution for the variable $y \in \cY$ given a linear projection $\rho(x) = \bf w^\top x$ of the input $x \in \cX$. $\rho$ is called the predictor. The distribution of $y|\rho(x)$ is chosen to be in the exponential family. This includes, for example, the Bernoulli, the Gaussian and the Poisson distribution that are used to deal with binary, continuous and count data respectively. GLMs are very popular tools due to their good estimation properties (the optimisation problem is convex and iterative estimation procedures converges rapidly), its ease of application and the fact that it can accommodate for many types of data (binary, categorical, continuous) both for \input{} and \output{}. 
The magnitude of the weight $\bf w$ is interpreted as indicating the impact of the corresponding factor on the observed data, with a value of $0$ indicating an absence of impact. \\

However, GLMs are intrinsically limited by their underlying assumption that the predictor linearly depends on the \inputs. In most situations, we expect regressors to have some \nonlinear impact onto the neural activity or behavior.
Generalized Additive Models (GAMs) are a non-linear extension of GLMs where the linear predictor is replaced by an additive predictor $\rho(x) = \sum_k f_k(x)$ where $f_k$ are functions from $\cX \to \cY$. Each $f_k$ usually depends on one or a subset of dimensions $s_k \subset [1..dim(\cX)]$ of $x$ (we will make this implicit in the rest of the article by using $f_k(x)$ instead of $f_k(x_{s_k})$). Functions $f_k$ need to be constrained to have some quantifiable form of regularity (e.g. to be smooth), both to make the problem identifiable and to capture a priori assumptions about these functions.  Additivity in GAMs introduces another form of identifiability, as functions are only defined up to an offset (i.e. replacing $f_1$ by $f_1 + \lambda$ and $f_2$ by $f_2 - \lambda$ for any $\lambda \in \mathbb R$ does not change the model). For this reason, it is convenient to add one constraint on each function (i.e. $f_i(0)=0$) and model the shift factor as an extra parameter $c \in \mathbb R$ to be estimated: $\rho(x) = \sum_k f_k(x) + c$.
\\

GAMs release the linearity constraint from GLMs that regressors are mapped linearly onto the predictor. However, one may want to go beyond the additivity hypothesis and allow for \nonlinear interactions across the variables in the regressor.

\subsection{Generalized Unrestricted Model (\gum)}
\begin{figure}
\centering
  \includegraphics[width=.7\linewidth]{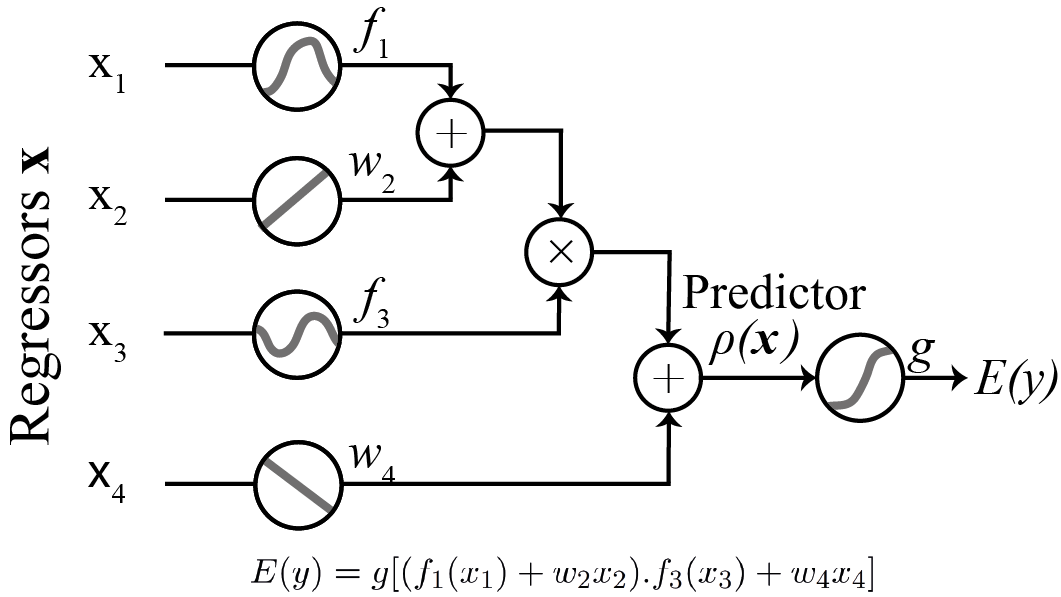}
  \caption{Example of \gum. Here 4 \inputs{} $(x_1, x_2, x_3, x_4)$ are combined to generate a prediction for \output{} $y$. Each \input{} is first passed through a linear function $w_i x_i$ or \nonlinear function $f_i(x_i)$. Then \inputs{} are combined with a set of additions and multiplications to yield the predictor $\rho(\vx)$, in this example $\rho(\vx)=(f_1(x_1)+w_2x_2)f_3(x_3) + w_4 x_4$. Finally, the expectation for \output{} $E(y)$ is computed by passing the predictor through a fixed function $g$ (the inverse link function). Inference corresponds to estimating the set of function $(\bf w, \bf f)$ based on a dataset $(\vX,\vy)$.}
  \label{fig:gum_architecture}
\end{figure}

One way to achieve is to extend GAMs and \emph{add} terms that are multiplications of functions to be learned. For example, we would like to capture models such as $\rho(x) = f_1(x)f_2(x) + c$, $\rho(x) = (f_1(x) + f_2(x))f_3(x) + c$, $\rho(x) = f_1(x)f_2(x) + f_3(x) + c$ or $\rho(x) = \sum_i f_1(x_i)f_2(x_i)f_3(x_i) + c$.
We define a Generalized Unrestricted Model (\gum) as a regression model composed of the following features: a set of functions or components $f_k$ defined over the input space; a predictor function $\rho(\vx)$ composed by summations and multiplications of the components; an observation model $y|\rho$ in the exponential family (Figure \ref{fig:gum_architecture}).\\

The predictor can be constructed recursively using summations and multiplications over the functions. We will study here a relatively general form, where the predictor is a sum of products of sums of GAM predictors (Figure \ref{fig:gum_architecture}): 
\begin{equation} \label{eq:gum}
\rho(x)  = \sum_\c\prod_{\d \leq D_\c} \big(\sum_l \f(x)\big)
\end{equation}

All of the predictors presented in the previous section can be expressed in such form. As for GLMs and GAMs, \gum{} regression consists in estimating the components $f_k$ given a dataset of \input{} $\vX = (\vx^{(1)},\dots, \vx^{(N)})$ and a corresponding output $\vy = (y^{(1)},\dots, y^{(N)})$.
We refer to a factor as a sum of functions $\big(\sum_l \f(x)\big)$, and to a block as the product of factors. The predictor is thus built as a sum of blocks. The dimensionality of each block $D_\c$ is the number of products within the block, i.e. the dimension of the multilinearity. A GAM can be seen as a \gum{} where all the blocks have dimensionality 1.

\subsection{Identifiability of \gum{}s}

Using such a large class of models requires great care to ensure model identifiability. The summation of functions induces a shifting degeneracy, while the multiplication of functions induces a scaling degeneracy: $f_1(x)f_2(x) = \frac{f_1(x)}{\lambda}(\lambda f_2(x))$\\ for any $\lambda \neq 0$. A general solution to this problem is to constrain all the functions to verify $\f(x_0)=0$ for some given $x_0$ and to then add offsets $\offset \in \mathbb R$ to the model, defining now $\rho(x)  = \sum_i\prod_j\big(\sum_l \f(x) + \offset \big) + c_0$. Equivalently, functions can be constrained to have mean 1 over a certain set of values We add some further constraints on the offset depending on the model itself (see details in Appendix \ref{appendix_constraints}). The examples of models provided above will write as: $\rho(x) = (f_1(x)+c_{11})(f_2(x)+1) + c_0$, $\rho(x) = (f_1(x) + f_2(x) + c_{11})(f_3(x)+1) + c_0$ or $\rho(x) = (f_1(x)+c_{11})(f_2(x)+1) + f_3(x) + c_0$. \\

\subsection{Bayesian treatment of \gum{}s}

Recent progress in Bayesian statistics allows to derive algorithms to perform inference in such models that are both accurate and scalable, meaning that they can be efficiently applied to the large datasets produced in neuroscience today. Adopting the framework of probabilistic modelling \cite{mackay2003information, bishop2006pattern}, we frame the model fitting task as probabilistic inference and learning problems. To do so we treat the functions and parameters of the model as latent variables and encode our \emph{a priori} assumptions about these in the form of distributions (using Gaussian processes as priors over functions \cite{rasmussen2005}). Hyperparameters control the statistical properties of the functions, for example their smoothness or perioditicity. Inference refers to estimating the functions $f$ and offsets $c$ of the model for a given value of the hyperparameters, while learning refers to estimating the hyperparameters. Due to the model structure, the inference problem is intractable and we resort to an approximate Bayesian inference technique called variational inference \cite{blei2017variational}.\\

The remainder of this paper is structured as follows.  Section 2 reviews the relevant background related to Gaussian process inference and previous work in sparse  approximations to GP models.  In Section 3 the core methodology for GUMs is presented.  Implementation  details  and  computational complexity of the method are covered in Section 4.  Section 5 is dedicated to a set of illustrative  toy  examples  and  a  number  of  empirical  experiments,  where  practical  aspects of GUM inference are demonstrated.  Finally, we conclude the paper with a discussion in Section 7.

\section{Methods}

\subsection{Probabilistic modelling and inference}

\subsubsection{General framework}

We propose to tackle the problem of learning the functions and parameters of our regression models as probabilistic inference problems.
To do so we start by defining a joint distribution over the \output{} $\vy$, and the parameters $\vtheta$ given the \inputs{}{} $\vx$, with density: $p(\vy, \vtheta | \vx)  = p(\vy | \rho_{\vtheta}(\vx)) p(\vtheta )$, where the first term is the exponential family observation model inherited from the GLMs and $p(\vtheta)$ is an \emph{a priori} distribution over the parameters capturing our statistical beliefs or assumptions about their values. In the case of {\gum}s, parameters correspond to functions and constraints, $\vtheta = \{f_k, c_k\}$. Both the prior and likelihood may be parameterised by hyperparameters $\gamma$. As is the case in most regression models, we will assume that conditioned on the parameters, the observations are statistically independent, i.e. the likelihood factorizes across the data points:  $p(\vy|\vtheta) = \prod_n p(y\n| \vtheta_n)$, where $\vtheta_n \subseteq \vtheta$ is the subset of parameters on which observation $y\n$ depends.\\

We then propose to treat the inference problem as that of computing the \emph{posterior} distribution over the parameters which is defined as their conditional distribution given the observed data $p(\vtheta|\vy)$. The posterior density can be expressed using Bayes' rule as $p(\vtheta|\vy) = \frac{p(\vy|\vtheta)p(\vtheta)}{p(\vy)}$. In this expression, $p(\vy)= \int  p(\vy, \vtheta) d\vtheta$ is the marginal likelihood which is commonly used as a objective to select the value of the hyperparameters $\vgamma$).\\

\subsubsection{Gaussian processes: distributions over functions}

Gaussian processes are commonly used as prior over functions \cite{rasmussen2005} because they can flexibly constrain the space of acceptable solutions in non-linear regression problems. Formally, 
Gaussian  processes  are  infinite  collections  of  random  variables, any finite subset of which follows a multivariate normal (MVN) distribution. They are defined by a mean function $m$ and covariance function $k$. A sample from a GP defined on an index set $X$ is a function on the domain $\cX$.  Given a list of points $X \in \cX^N$ and a GP sample $f\sim GP(m,k)$, the vector of function evaluations $f(X)$ is an associated MVN random  variable  such  that $f(X) \sim \NN(m(X),K(X,X))$, where $m(X)$ is a vector of mean function evaluations and $K(X,X)$ is a matrix of all pairwise covariance function evaluations ($K(X,X)_{ij}=k(x_i, x_j)$). The covariance function may depend on some hyperparameters $\vgamma$ (the mean function can be too, although in most applications it is taken to be the zero function). For example, the classical Squared Exponential (SE) covariance function defines covariance for real-valued data $\cX = \mathbb R^d$ based on two hyperparameters: length scale $\ell$ (which parametrises the expected smoothness of the function) and variance $\beta^2$ (which parameterises the expected magnitude of the function) ($k_{SE}(x,x')= \beta^2 e^{-|x-x'|^2/2 \ell^2}$). As such, the covariance and mean functions define soft constraints over the functions $f(X)$ that fulfill two goals: make the functions $f(X)$ identifiable, and formalizing assumptions about the possible forms that $f(X)$ may take (i.e. its smoothness).

\subsubsection{Variational inference}
\label{sec:vi}

Evaluating the posterior density $p(\vtheta|\vy)$ and the marginal likelihood $p(\vy)$ is in general intractable for a \gum. We therefore resort to perform approximate inference and focus on variational inference methods \cite{blei2017variational}. Variational inference turns the inference problem into an optimisation problem by introducing a variational distribution over the parameters $q(\vtheta)$ and maximising a lower bound $\cL(q)$ to the log marginal evidence $\log\,p(\vy)$.
This lower bound is derived using Jensen's inequality :
\begin{align}  
\label{eq:elbo}
\log\,p(\vy) = \log \int d\vtheta \frac{p(\vy, \vtheta)}{q(\vtheta)} q(\vtheta)  \geq \int q(\vtheta) \log \frac{p(\vy, \vtheta)}{q(\vtheta)} = \cL(q)
\end{align}
The lower bound can be rewritten as 
\begin{align}
\label{eq:elbo_impl}
\cL(q) = \sum_n \EE_{q(\vtheta_n)} \log\,p(y_n|\vtheta_n) 
- KL[q(\vtheta)|p(\vtheta)],
\end{align}
which is the form favored for actual implementations. The left-hand terms in the sum are the \emph{variational expectations} and require computing an expectation under the marginal distributions $q(\vtheta_n)$.\\

The gap in the inequality in Equation \ref{eq:elbo} can be shown to be $\log\,p(\vy) - \cL(q) = KL[q(\vtheta)|p(\vtheta|\vy)]$, where KL denotes the Kullback-Leibler divergence between distributions. Hence, as  $q(\vtheta)$ gets closer to the true posterior $p(\vtheta|\vy)$, the bound gets tighter. In practice, one chooses the class of distribution $\cQ$ such that optimising $\cL(q)$ for $q$ is tractable. Once optimised, the optimal variational distribution $q^{*}(\vtheta)$ provides an approximation to the posterior and the bound $\cL(q^{*})$ provides an approximation to the log marginal likelihood.\\

The choice of the class $\cQ$ is driven by two opposite goals: $\cQ$ must be rich enough so that at the optimum, $q^*$ captures most important features of $p(\vtheta|\vy)$, but simple enough that $\cL(q)$ is computationally efficient to evaluate for all $q \in \cQ$. When a finite dimensional parameter $\vtheta$ is endowed a prior distribution which is a multivariate normal (MVN) distribution, a convenient choice for $\cQ$ is the class of MVN distributions parameterised by a mean vection $\vmu_q$ and a covariance $\vSigma_q$, i.e. $q(\vtheta)=\NN(\vtheta; \vmu_q, \vSigma_q)$ \cite{challis2013gaussian}. In that case, computing the marginals $q(\vtheta_n)$ is straightforward and the KL between two MVNs has a simple expression. The evaluation of the variational expectations in Equation \ref{eq:elbo_impl} can be evaluated in closed form or approximated using Gaussian quadrature or Monte Carlo methods \cite{hensman2013gaussian}.

When working with functions $f(\cdot)$ and using Gaussian Processes  - infinite dimensional objects - as priors, an efficient and scalable way to parameterise the variational distribution as a finite dimensional Gaussian Process expressed at some input $z\in \cX^M$ as follows: 
\[q(f(\cdot), \vf) = p(f(\cdot)|f(\vz)=\vf)q(\vf),\]

where $\vf= f(z)$, $q(\vf)=\NN(\vf; \vmu_{\vf}, \vSigma_{\vf})$ is a MVN distribution and $p(f(\cdot)|f(\vz))$ is the conditional prior process. $q(\vf)$ can be interpreted as an approximate marginal posterior on $f(\vz)$.
Choosing $z=x$ is the classical variational treatment of GPs but leads to a computational cost to evaluate $\cL(q)$ that scales cubically with $N$, making the use of variational inference prohibitively expensive for large datasets.
Choosing $z\in \cX^M$ to be a set of pseudo-inputs (or \textit{inducing points}) of size $M$ leads to the so-called  sparse variational approach \cite{titsias2009variational, hensman2013gaussian, matthews2016sparse, bauer2016understanding} whose $\bigO(NM^2 + M^3)$ complexity allows to scale variational inference with GPs to large datasets.\\

In order to enforce the necessary constraints that make the model identifiable, it is possible to force the posterior processes to have a predictive mean at an input $x_c$ to be equal to $y_c$. To do so we adjust the variational mean by a scaled unit vector $\vmu_{\vf} \leftarrow \vmu_{\vf} + \delta \bf{1}$ such that $\EE_q[f(x_c)]=y_c$, where $\delta = \frac{y_c - \EE_p[f(x_c)|f(\vz)=\vmu_{\vf}]}{\EE_p[f(x_c)|f(\vz)=\bf{1}]}$. This is the method we use in our examples.

\subsubsection{Laplace approximation}

The Laplace approximation provides an alternative MVN approximation $q(\vtheta) = \NN(\vtheta;\vmu_q, \vSigma_q)$ to the posterior $p(\vtheta|\vy)$. This approximation is derived from a Taylor expansion of the log-joint density $\log p(\vtheta, \vy)$ taken at the maximum a posteriori parameters $\vtheta\map= arg\max_{\vtheta} p(\vtheta,\vy)$:
\beq
 \log p(\vtheta, \vy) \approx a + b (\vtheta-\vtheta\map) - \frac 1 2(\vtheta-\vtheta\map)^T H(\vtheta-\vtheta\map) 
\eeq 
where $a = \log p(\vtheta\map, \vy)$, $b = \nabla_{\vtheta} \log p(\vtheta,\vy) |_{\vtheta=\vtheta\map} = 0$ by definition of $\vtheta\map$, and $H = - \nabla \nabla_{\vtheta} \log p(\vtheta,\vy) |_{\vtheta=\vtheta\map}$. $H$ is a definite-positive matrix since it is the opposite Hessian of a function evaluated at its maximum. The expansion leads to
\beq
   p(\vtheta| \vy) \propto p(\vtheta, \vy) \approx e^{a}\exp\left(- \frac 1 2(\vtheta-\vtheta\map)^T H (\vtheta-\vtheta\map)\right)
\eeq

The constant of proportionality is obtained from the normalization constraint  $\int p(\vtheta| \vy) d \vtheta = 1$, yielding the MVN form $p(\vtheta| \vy)= \NN(\vtheta;\vmu_q, \vSigma_q)$ with  $\vmu_q = \vtheta\map$ and $\vSigma_q = H^{-1} = [-\nabla \nabla_{\vtheta} \log p(\vtheta,\vy) |_{\vtheta=\vtheta\map}]^{-1}$.  \\
The Laplace approximation is a fairly simple procedure, as it only requires to find the maximum a posteriori parameters $\vtheta\map$, and to then compute the Hessian of the log-joint probability evaluated at $\vtheta\map$. 
The approximation is increasingly better as the sample size $n$ increases, as the true posterior becomes closer to a MVN distribution \cite{bishop2006pattern}. However the computational complexity of the method scales cubically with the number of data points.

\subsection{Approximate inference for \gum s}

We present two different approximate methods to treat GUMs: Laplace approximation and sparse variational inference. For each method, we describe first how inference can be performed, i.e. how the posterior over the functions $f$ and offsets $c$ can be approximated from a dataset, assuming fixed values of the hyperparameters for the different GPs. Then, for each method, we describe how these hyperparameters can also be learned from the dataset.

\subsubsection{Laplace approximation for \gum s}

\paragraph*{Inference}

For classical GP classification, the Laplace approximation follows by recognizing that the problem is formally equivalent to a Bayesian GLM, where the parameters are the value of the GP at data points $\vf = f(\vx)$, the prior covariance is given by the GP evaluated at data points $K(\vx,\vx)$ and the design matrix is identity \cite{rasmussen2005}. Then the MAP solution $\vf\map$ can be found iteratively using Newton-Raphson updates (the joint density is convex), and the Hessian can be evaluated analytically. Computing the Laplace approximation for GUMs follows a similar path. The derivation is quite lengthy, so we summarize here the main steps (details are provided in Appendix \ref{appendix_laplace}). \\ First, a \gum{} can be turned into a Bayesian formulation of a generalized multilinear model \cite{Shi2014, ChristoforosChristoforou2010}, as multiplications of functions yield multilinear interactions of the parameters $\vf_k=f_k(\vx)$. The constraints on $f_k$ added to remove identifiability problems lead to removing one free parameter for each parameter set $\vf_k$. The MAP solution is then found by iteratively performing Newton-Raphson update on each dimension while leaving others parameters unchanged. For example for a \gum{} $\rho(x)= f_1(x)f_2(x)$, we update the value of $\vf_1$ while leaving $\vf_2$ unchanged, then we update $\vf_2$ while $\vf_1$ is unchanged, and loop until convergence. Each iteration increases the value of the joint density.
Finally, the covariance of the approximated posterior can be computed analytically by evaluating the Hessian of the log joint density at the MAP parameters.

\paragraph*{Hyperparameter fitting}

 For Laplace approximation we describe two different ways of fitting the hyperparameters $\hp$ for the prior covariance functions $\Kcd = \Kcd(\hp\ij)$: cross-validation and generalized expectation-maximisation \cite{Wood2011}. \\
 In cross-validation, we split the dataset between a training set and a test set (possibly multiple times, as in K-fold cross-validation). We infer the MAP estimate $\vtheta\map$ using the training set only, then compute the cross-validated log-likelihood (CVLL), i.e. the log-likelihood of the MAP parameters evaluated on the test set $\log p(\vy_{\text{test}}| \vtheta\map)$. The gradient of the CVLL over the hyperparameters can be calculated analytically (see Appendix \ref{Laplace_CVLL}). This allows to run a gradient ascent algorithm which iterates between computing the MAP estimates for given hyperparameters and then updating the hyperparameters to along the gradient to improve the CVLL score.\\
 
 In expectation-maximisation, we used the  approximated posterior $q(\vtheta)$ to update the lower bound on model evidence $\cL(q, \vgamma)=\int q(\vtheta) \log p(\vy|\vtheta, \vgamma) d\vtheta + const$. The algorithm iterates between the expectation step (inference using Laplace approximation) and the maximisation step where the lower bound is maximised with respect to the hyperparameters. Because both priors and posteriors are Gaussian, the lower bound can be expressed analytically and maximised using gradient ascent. Note however that, because the posterior is not exact but approximated, the lower bound is not guaranteed to increase at each expectation step.

\subsubsection{Sparse variational approximation for \gum s}

\paragraph*{Inference}

Variarional inference has been been used to learn GLMs \cite{nickisch2009vglm} and GAMs \cite{hui2019semiparametric, adam2016scalable}. In these settings, having multiple functions, we need to specify a variational posterior over the functions and scalar parameters $q(f_{1\dots K}, \vc)$. We follow \cite{adam2017structured} and assume it factorizes as
\[q(f_{1\dots K}, \vc) = q(\vf_{1\dots K}, \vc) \prod_k p(f_k|f_k(\vz_k) = \vf_k),\]
which means that posterior processes are only coupled through the inducing variables $\vf_{1\dots K} \in \RR^{\sum_k M_k}$ ($M_k$ is the number of inducing points for $f_k$). \\

We are left to characterize the finite density $q(\vf_{1\dots K}, \vc)$. We choose to parameterise it as a MVN distribution. A first option is to parameterise it as fully coupled MVN distribution, which would capture the posterior coupling but would also do so in a overparameterised fashion. The other extreme would consist in a mean field approximation across term $q(\vf_{1\dots K}, \vc) = \prod_k q(\vf_k)\prod_i q(c_i)$ which would under-estimate the posterior variances \cite{bishop2006pattern} and possibly bias learning \cite{turner2011}. \\

We propose a parameterisation that preserves coupling across functions:  $q(\vf_{1\dots K}, \vc) = q(\vf_{1\dots K})\prod_i q(c_i)$, where we let each factor be a MVN distribution. For each function we enforce $q(f_k(0))=0$ using the method described in Section~\ref{sec:vi}.

\paragraph*{Hyperparameter fitting}

When using variational inference,  $\cL(q)$ is used as a proxy to the log marginal likelihood and can be optimized with respect to the hyperparameters $\vgamma$ as well as with respect to $q$, an approach akin to Expectation Maximisation \cite{bishop2006pattern}. We here follow this approach: we iterate between maximizing $\cL(q)$ over $q$ with $\vgamma$ fixed (inference) and maximizing $\cL(q)$ over $\vgamma$ with $q$ fixed (learning).


\section{Results}

\subsection{Synthetic data}

We first tested the ability of the different algorithms to infer the correct form of functions from synthetic data. We generated Poisson observations $\vy$ ($E(y|\rho) = e^{\rho}$) from a simple GUM of the form $\rho(\vx) = f_1(x_1)f_2(x_2) + f_3(x_3)$ and then applied our algorithms to fit the functions $(f_1, f_2, f_3)$ on the dataset $(\vx, \vy)$.
We chose $f_1$ and $f_3$ to be functions of a closed interval $ [ 0,2]$ and $f_2$ to be periodic (of period $\pi$). This could correspond for example to regressing the spiking activity of a neuron in visual cortex against the properties of visual stimulus defined by its contrast and orientation: $f_1(x_1)$ would be the modulation of firing rate by contrast, $f_2(x_2)$ the orientation selectivity, and $f_3(x_3)$ could be the fluctuation of neural excitability across time (with $x_3$ being time).
More specifically, we defined $f_1(x_1)=\exp(x_1/2)-1$, $f_2(x_2)= 1+ \cos(2x_2 + \pi/3)$ and $f_3(x_3) = - \sin(x)$. 
For estimation of $f_1$ and $f_3$ we used the standard Squared Exponential kernel $K_{SE}(x,x')= \beta^2 \exp\big( - \frac{|x-x'|^2}{2 \ell^2} \big)$ with the value of the hyperparameters $\beta= 1$ and length scale $\ell= 0.1$. For $f_2$ we used the standard periodic kernel $K_{per}(x,x')= \beta^2 \exp \big( -2 \sin^2(\frac{\pi}{T}|x-x'|) / \ell^2   \big)$, with hyperparameters $\beta= 1$,  $\ell= \pi/20$ and period $T=\pi$. \\

\begin{figure}
\centering
  \includegraphics[width=.8\linewidth]{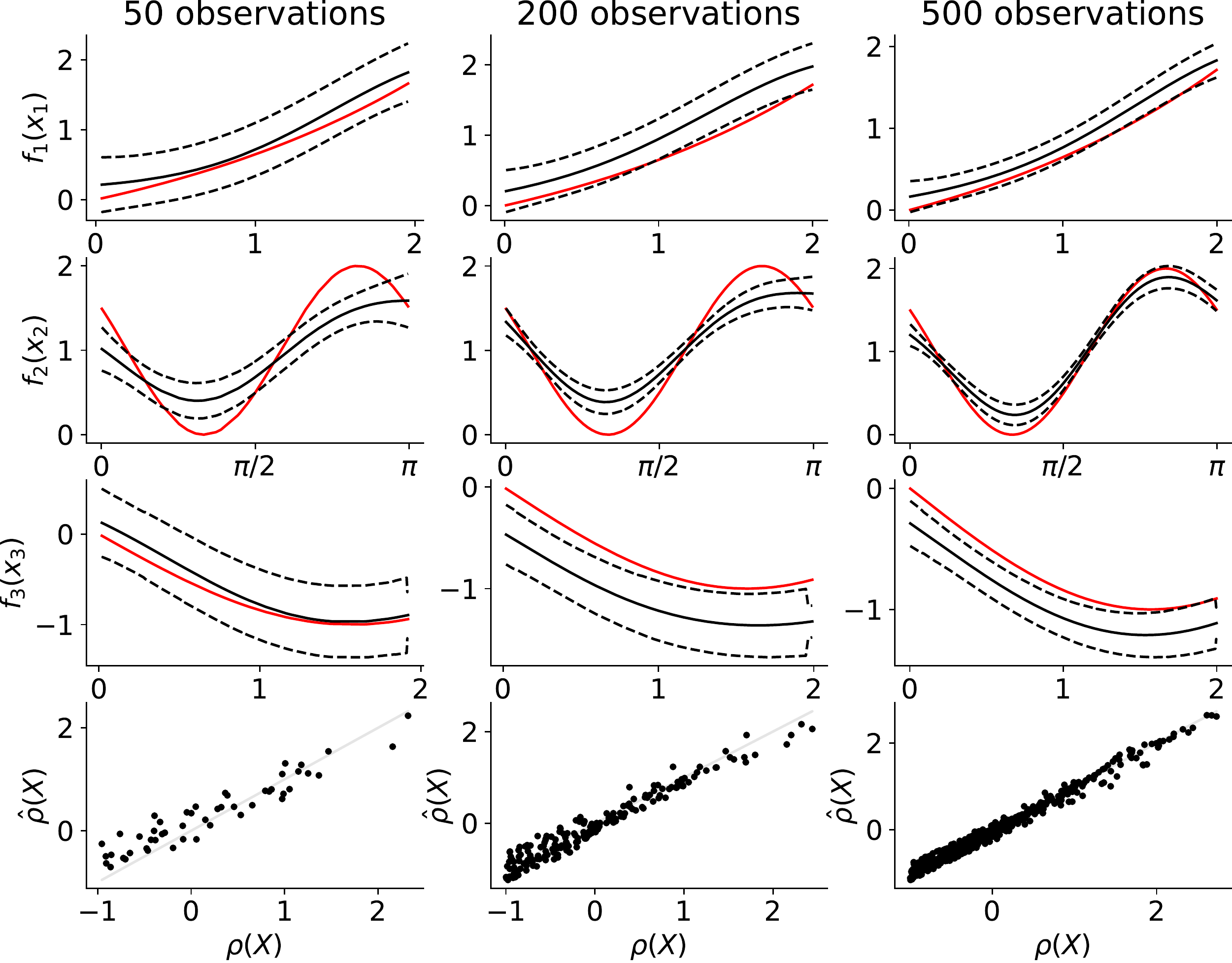}
  \caption{Estimated functions using Laplace method from \gum{} model with predictor $\rho(\vx) = f_1(x_1)f_2(x_2)+f_3(x_3)$. Posterior distribution for each function $f_i$ is a Gaussian Process $f_i \sim GP(m_i,k_i)$. Solid black lines depict the posterior mean function $m_i(x)$,  dotted lines depict the posterior standard error. Red lines depict the ground truth, i.e. the functions that were used to generate the dataset. Different columns represent different sizes for the dataset used to estimate the functions (50, 200 or 500 observations).}
  \label{fig:synthetic_functions}
\end{figure}

\begin{figure}
\centering
  \includegraphics[width=.8\linewidth]{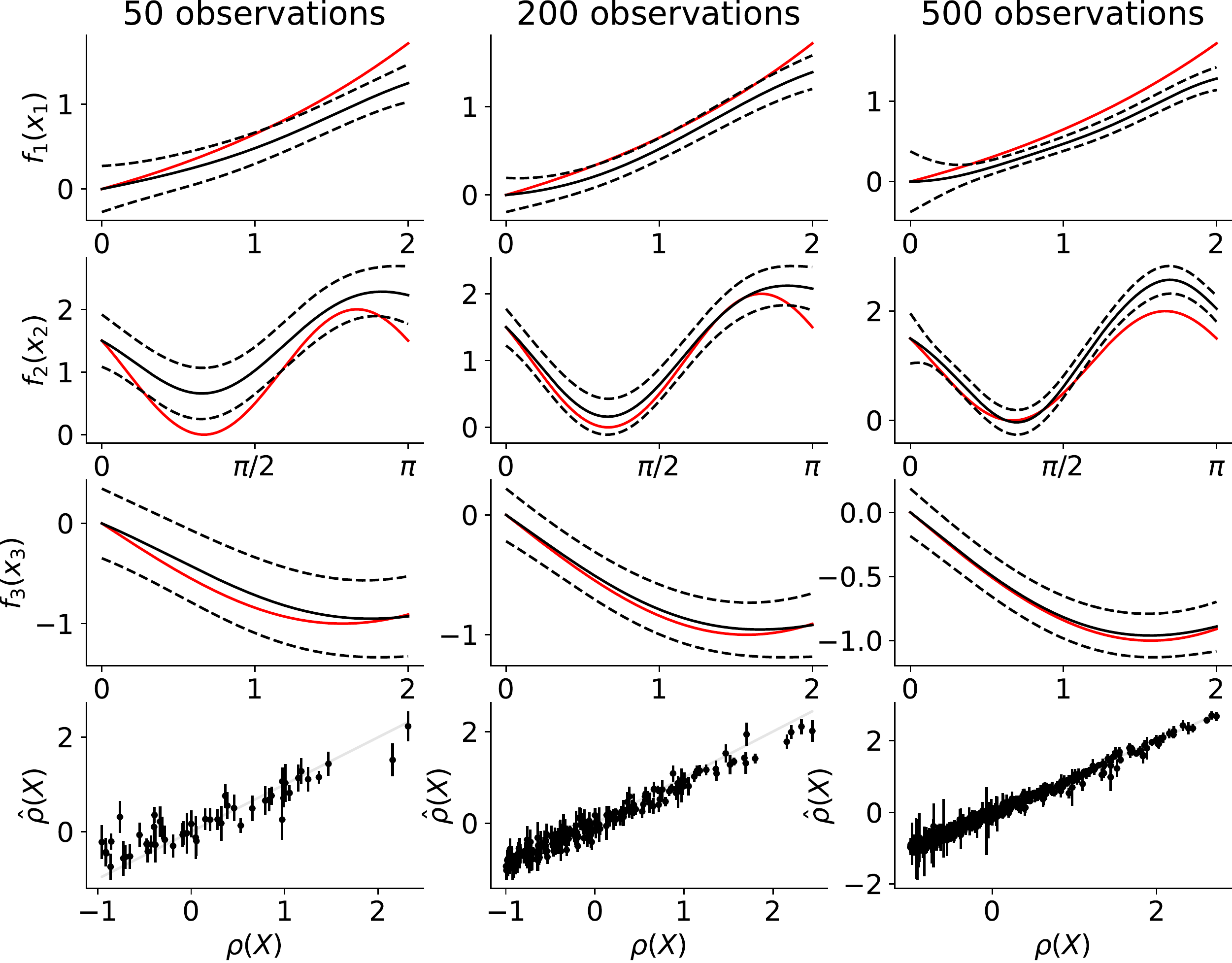}
  \caption{Estimated model using the variational inference for  \gum{} model with predictor $\rho(\vx) = f_1(x_1)f_2(x_2)+f_3(x_3)$, for same dataset as figure 2. Top three rows: posterior distribution for each function $f_i$ is a Gaussian Process $f_i \sim GP(m_i,k_i)$. Solid black lines depict the prior mean function $m_i(x)$,  dotted lines depict the posterior standard error. Red lines depict the ground truth, i.e. the functions that were used to generate the dataset. Different columns represent different sizes for the dataset used to estimate the functions (50, 200 or 500 observations). Bottom row: posterior predictive predictor $q(\rho)$ against ground truth.}
  \label{fig:vi_toy_results}
\end{figure}

\begin{figure}[p]
  \centering

  \includegraphics[width=.8\linewidth]{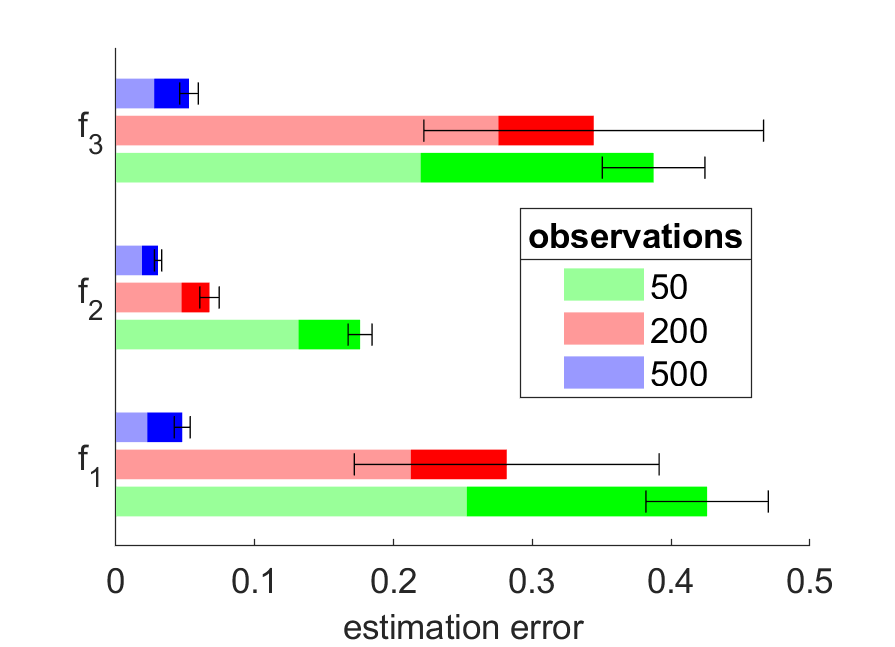}
  \caption{Estimation error for functions $f_1$, $f_2$ and $f_3$ in \gum{} with predictor $\rho(\vx) = f_1(x_1)f_2(x_2)+f_3(x_3)$, as a function of the number of observations. Values represent the average expected error over 30 repetitions of GUM inference (as in Figure \ref{fig:synthetic_functions}). Error bars represent the s.e.m. Lighter bar represents the proportion of the due due to error in posterior mean, while darker portion represents the proportion due to posterior variance (see \ref{sec:measuring_error} for details).
  }
  \label{fig:synthetic_error}
  \includegraphics[width=.6\linewidth]{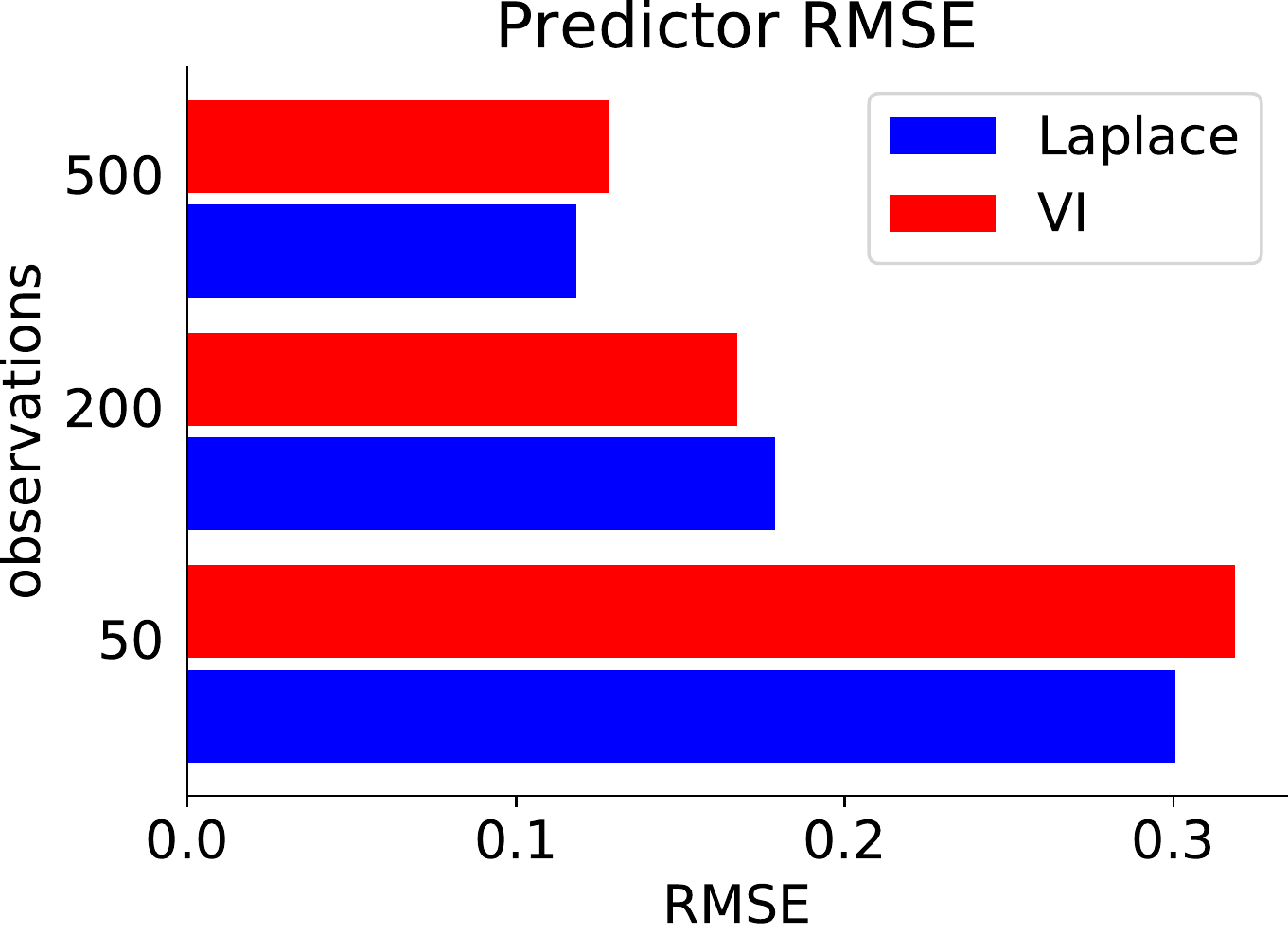}
  \caption{Root Mean Squared Error (RMSE) of the posterior predictor against ground truth for both the Laplace and the Variational inference methods.}
  \label{fig:predictor_error}
\end{figure}

We report the inferred functions and the resulting predictors for both methods in Figure \ref{fig:synthetic_functions} (Laplace method) and Figure \ref{fig:vi_toy_results} (sparse variational inference) which show the functions are correctly recovered, even for small sized. The reconstruction error is compared between the different sizes of the dataset ($N=50,200,500$ observations) in Figure~\ref{fig:synthetic_error}. Using priors over the functions implies that the estimator will be biased (towards zero), and that the bias will be larger for smaller datasets. In practice, however, such shrinking effect was very small for $N=200,500$, and was only pronounced for the smaller sample size ($N=50$) for the estimation of $f_2$. As expected, all three functions were more accurately estimated for larger sample sizes. 
Both methods achieve similar regression performance as measured by the root mean squared error (RMSE) on the predictor, which decreases as the number of observations grows (Figure \ref{fig:predictor_error}). In terms of computing however, the sparse variational method clearly outbeats the Laplace method for large sample size (from $N=500$).
For the Laplace method, we also report the reconstruction error between the true function $f_i$ and estimated function $\hat f_i$ as $err_i= <(f_i(x_i)-\hat f_i(x_i))^2>_{x_i}$.

\subsection{Experiments: psychophysical data - perceptual decision study}
The very flexible functional forms of {\gum}s makes them applicable to a wide range of problems in psychophysics, neuroscience and beyond. We illustrate {\gum}s on experimental data from a study using a classical evidence accumulation paradigm \cite{Gold2007}.

A classical problem in cognitive neuroscience is to understand how humans integrate evidence from multiple sources of information to make decisions \cite{Gold2007}. One embodiment of this problem is the study of how this integration happens across time given a temporal sequence of stimuli, such as oriented gratings \cite{Wyart2012a}. 
In such paradigms, subjects report through a binary decision the value of a particular statistic of a stimulus feature across the sequence (e.g. whether gratings are mostly tilted clockwise or counterclockwise, figure \ref{fig:grating_paradigm}). 

To model the contribution of the different visually presented gratings $x_{tk}$ (where $t$ denote trial number and $k$ the index of the grating in the sequence) to the behavioral response $y_t$, a binomial GLM (i.e. logistic or probit regression) is frequently used. Such a model has the form: $p(y_t=1)=\sigma(\rho)$ and $\rho=\sum_k w_k z_{tk} + c$, where $w_k$ is the weight of the stimulus in position $k$, $c$ is the lateral bias towards one response and $z_{tk} = f(x_{tk})$ is a (possibly \nonlinear) transformation defined by the normative framework (formally, $z_{tk} = \log \frac{p(x_{tk}|\hat y =1)}{p(x_{tk}|\hat y =0)}$). In the grating task task, $z_{tk}=\cos( x_{tk}-\theta_{\text{ref}})$.

\subsubsection{Learning a mapping from stimuli to evidence}
The {\gum\,} framework allows to extend the types of models we can fit beyond this standard GLM, and capture rich interactions between factors to explain behavior.
Notably, the mapping $f$ from sensory to perceptual evidence may depart from the normative standpoint. The particular shape of the mapping may depend on how orientation is encoded by neural populations in the visual cortex, or the way the task is presented to the participants. We can formulate a \gum{} model where such mapping $f$ is learned from data rather than defined \textit{a priori}: $\rho(\vx)=\sum_k w_k f(x_{tk}) + w_0$. This model relies on the multiplicative interaction between the stimulus weights $w_k$ and the \nonlinear mapping $f$, one defining feature of \gum s. 
This \gum{} is related to a strictly additive model (GAM) $\rho(\vx) =\sum_k f_k(x_{tk}) + w_0$ where one stimulus mapping $f_k$ is defined for each position in the sequence. However this latter model ignores the fact that all stimuli are similar in nature and processed by the same sensory areas, so that the mapping should be conserved up to a scaling factor. Adding the scaling constraint $f_k(x) =w_k f(x)$ gives the GUM equation,  a model with better interpretability and with less parameters (a single mapping function to be fitted), so that it can be inferred precisely with less observations.\\

\begin{figure}
  \includegraphics[width=\linewidth]{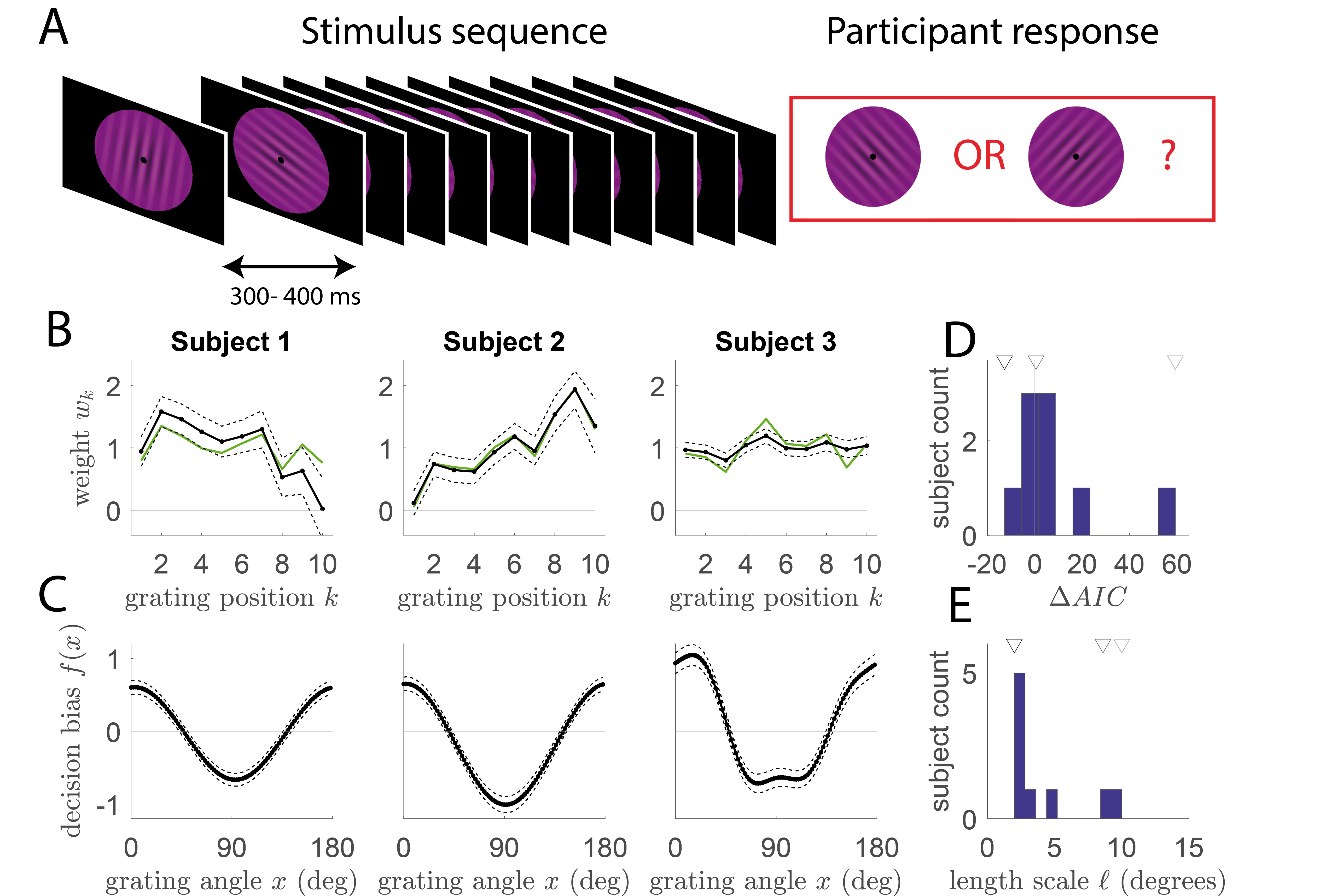}
  
  \begin{subfigure}{0em}
        \phantomsubcaption{}
        \label{fig:grating_paradigm}
    \end{subfigure}
    \begin{subfigure}{0em}
        \phantomsubcaption{}
        \label{fig:grating_kernel}
    \end{subfigure}
    \begin{subfigure}{0em}
        \phantomsubcaption{}
        \label{fig:grating_mapping}
    \end{subfigure}
      \begin{subfigure}{0em}
        \phantomsubcaption{}
        \label{fig:grating_lengthscale}
    \end{subfigure}
    \begin{subfigure}{0em}
        \phantomsubcaption{}
        \label{fig:grating_AIC}
    \end{subfigure}

  \caption{Application of \gum{} to analysis of human psychophysics experiment
  \textbf A. Behavioral paradigm. In each trial, the participant viewed a sequence of 5-10 visual gratings with a certain orientation, interspersed with a 300-400 ms interval. Subjects had to report at the end of the sequence whether the average orientation of the gratings was more tilted clockwise or counter-clockwise.
  \textbf B. Psychophysical kernels recovered from the \gum{} analysis for 3 exemplar subjects. The kernel represents the weight of each grating $w_k$ as a function of its position $k$ in the stimulus sequence. Full black line represents the posterior mean, and dotted black lines the standard deviation, obtained using the Laplace method. The green line represents the weights obtained from the standard GLM analysis.
  \textbf C. Perceptual mapping for each subject, i.e. the decision update $f(x)$ for each grating as a function of its orientation $x$ relative to the reference orientation (orientation tilted clockwise, i.e. 45 degrees). Positive (resp. negative) values indicate that the grating biases the decision towards the 'tilted clockwise' (resp. 'tilted counter-clockwise') response. Perceptual mapping were estimated from \gum{} model, legend as in B.
  \textbf D. Histogram of difference in Akaike Information Criterion ($\Delta AIC$) between the \gum{} and simpler GLM, for all 9 participants. Positive values indicate that the \gum{} is favored, negative that that the GLM is favored. Triangles above indicate values for the 3 subjects of panels B-C (black: subject 1; dark grey: subject 2; light grey: subject 3). 
  \textbf E. Histogram of fitted values for hyperparameter $\ell$ that defines the expected length scale of mapping $f$.
  } 
  \label{fig:grating}
\end{figure}

We estimated the \gum{} model above on choice data from 9 participants that each performed 480 trials. We used a GP with periodic covariance function as a prior for $f$, and the function and its hyperparameters were estimated using the Laplace method with cross-validation. To handle identifiability problem, we constrained the weights to be on average 1 ($\frac 1 N \sum_{k=1}^N w_k=1$). Results are presented for three subjects in Figure \ref{fig:grating_kernel}-\ref{fig:grating_mapping}. First, different subjects displayed different  psychophysical kernels, i.e. different profiles of grating weight $w_k$. While subject 1 assigned more weight to gratings presented early in the sequence (the so-called \textit{primacy effect}), subject 2 assigned more weight to gratings showed late (\textit{recency effect}), and subject 3 displayed more or less equal weighting for all gratings. These patterns obtained from the \gum{} matched the profiles obtained from the more traditional GLM analysis.
More importantly, the \gum{} analysis permitted to recover for each subject how each grating was mapped on the decision space based on its angle, i.e. the decision mapping $f(x)$. the mapping of subject 1 is very similar to the cosine function predicted by the normative approach: gratings with relative angle of 0 (i.e. perfectly aligned with the reference grating) provided maximal bias in favor of the associated choice (rightward response), while gratings with relative angle of 90 degrees (i.e. perpendicular to the reference grating) provided maximal bias in favor of the alternative choice (leftward response). The mapping for subject 2 looked similar but with a vertical offset: leftward-tilted gratings  provided more bias towards left response than rightward-tilted did towards right response. Finally, mapping for subject 3 showed a much more abrupt transition from angles biasing the decision towards the left to angles biasing towards the right response. This is more consistent with a subject that simply categorizes the gratings as being tilted leftwards or rightwards and bases its decision based on the counts for each category, disregarding the precise angular distance of each grating to the references. It should be noted that GP always enforces a degree of smoothness to the recovered function, so that a step function could not be inferred with a finite dataset.\\
 We performed a model comparison to test, for each participant, whether the \gum{} provided a better account of the behavioral data than the simpler GLM model. This analysis shows whether using a flexible mapping instead of the fixed normative one (cosine function) improves the model. We used the Akaike Information Criterion, that corrects the approximate marginal evidence $\cL(q,\vgamma)$ with the number of hyperparameters $p$ ($AIC = 2p-2 \log \cL(q,\vgamma)$). The results were in agreement with what we observe for individual mappings (figure \ref{fig:grating_AIC}). For subject 1, whose mapping was very similar to the normative one, the GLM was favored. For subject 2 and 3, whose mapping differed from the normative one, the \gum{} was favored.  Finally, the fitted values of the hyperparameters provides an information about the degree of smoothness of the mapping that provided the best account of the data. In particular, the values of the length scale for the squared exponential was 4.1 $\pm$ 3.1 degrees (average $\pm$ std across participants) (figure \ref{fig:grating_lengthscale}).

\section{Discussion}
Here we have presented a novel class of regression models, Generalized Unrestricted Models or \gum s, that allows to capture nonlinear mapping for each \input, as well as additional and multiplicative interactions between these \inputs. We propose a Bayesian treatment of {\gum}s using the framework of Gaussian Processes, that allow to define distributions over functions with interpretable properties. Moreover, \gum s allow to perform regression for many different data type (including binary variable, categorical, real, periodic), making it an extremely versatile analysis method.

We have shown two different algorithms to learn {\gum}s from experimental data: the Laplace method and the sparse variational approach, which scales better for larger dataset. A \gum{} analysis on synthetic data showed that both methods allowed to recover mappings with low estimation error, even for small datasets. However, we strongly advise to run parameter recovery analysis for any new \gum{} problem. Indeed the estimation error will largely depend on the class of the model and the size of the dataset. For some classes of models, the identifiability may be poor.
Using multiple initial values for the estimation procedure is key to avoid finding local solutions to the estimation problem, but may not always correct this identifiability issue.\\
There is a long history of using regression analyses that capture multiplicative interactions between \inputs. ANOVA is routinely used to capture these interactions for categorical \inputs{} and continuous \output. This is equivalent to defining a GLM that includes multiplication of the \inputs{} into the design matrix (which is the method of choice for non-normally distributed \output). For continuous \inputs, previous studies have looked at extensions of the GLM and other regression models to include bilinear or multilinear terms \cite{ChristoforosChristoforou2010, DeFalguerolles2012,Ahrens2008, Shi2014, Dyrholm2007}. As in the Laplace method, estimation techniques for such models rely on alternatively updating the weights associated to one dimension while leaving the other weights fixed. However those regression models did not capture multiplication of  non-linearly transformed \inputs{} (i.e. $f_1(x)f_2(x)$). One exception is the work of Ahrens and colleagues who built multilinear models of neural spiking activity  \cite{Ahrens2008}, where each neuron firing rate $r(t)$ is based on spatio-temporal filtering of the acoustic stimuli $S(f,t)$ that is separable in time delay $d$ and frequency $f$, i.e. $E(r(t)) = \sum_{f,d} f_f(f)f_t(d)S(f,t-d)$. 
\gum s are also related to models designed to infer the latent dynamics of underlying low-dimensional factors from simultaneous spike recordings, such as Gaussian Process Factor Analysis (GPFA)\cite{Yu2009a} or variational Latent Gaussian Process (vLGP)\cite{Zhao2017}. For example, the predictor in vLGP for spike count for neuron $n$ at time $t$ is built as : $\rho_{nt} = \sum_k \alpha_{nk} f_k(t) + \sum_u \beta_{nu} y_{n,t-u} + c_n$, where $f_k$ is a collection of latent processes modelled as GPs, $\alpha_{nk}$ represent the weight of latent process $k$ onto each neuron $n$, $\boldsymbol{\beta}$ represent the impact of spike history onto neuron firing, and $c_n$ sets the baseline firing rate for neuron $n$. Spike count $y_{nt}$ is taken from a Poisson distribution with rate $\exp(\rho_{nt})$. This is one possible functional shape of a \gum. One small difference in treatment however is that weights $\boldsymbol{\alpha}$ and $\boldsymbol{\beta}$ were modelled as hyperparameters rather than latent processes themselves, in other words their solution did not model uncertainty about weight estimation.\\

In essence, the \gum{} framework expands the catalogue of models that can be estimated from data by adding multilinearity on top of nonlinear mappings. While it can be used as a purely predictive tool for machine learning applications, its primary development is for inference problems, where we are interested in estimating the nature of the influence of \inputs{} over a certain \output. The versatility of the tool, rather than simply expanding the space of possible predictive models, is meant to be at the service of a research question where interpretability is essential. We have provided an example here for analysis of behavioral data. In this example, the interaction between the functions of grating position and grating angle found a natural interpretation: the weights for grating position $w_k$ represent the impact of the grating depending on its position in the sequence, while the mapping from grating angle $f(x)$ represent how each grating biases the decision in favor of one choice or the other depending on its angle. We believe that many scientific questions in neuroscience and beyond could be explored using this new tool, for example: assessing decomposable spectro-temporal receptive fields from neural recordings \cite{Ahrens2008}; assessing complex cross-frequency coupling in neural signals \cite{Nadalin2019}; assessing how an evoked potential in EEG can be modulated parametrically by an experimental factor \cite{Ehinger2018}; etc. We are currently working on a toolbox to make this new versatile regression tool publicly available for analysis of neural and behavioral datasets.

\subsubsection*{Acknowledgements}
This research was supported by the Spanish Ministry of Economy and Competitiveness together with the European Regional Development Fund (PSI2015-74644-JIN and RYC-2017-23231 to A.H.). The authors would like to thank Isis Albareda for her help with the acquisition of behavioral data, V. Wyart for sharing code for the coding of the experimental paradigm, as well as J.Pillow and M.Aoi for fruitful discussions about the GP framework. 

\bibliographystyle{unsrt}
\bibliography{main}

\begin{thebibliography}{10}

\bibitem{mccullagh1989}
P.~McCullagh and J.~A. Nelder.
\newblock {\em Generalized Linear Models}.
\newblock Chapman \& Hall / CRC, 1989.

\bibitem{mackay2003information}
David~JC MacKay and David~JC Mac~Kay.
\newblock {\em Information theory, inference and learning algorithms}.
\newblock Cambridge university press, 2003.

\bibitem{bishop2006pattern}
Christopher~M Bishop.
\newblock {\em Pattern recognition and machine learning}.
\newblock Springer Science+ Business Media, 2006.

\bibitem{rasmussen2005}
Carl~Edward Rasmussen and Christopher K.~I. Williams.
\newblock {\em Gaussian Processes for Machine Learning (Adaptive Computation
  and Machine Learning)}.
\newblock The MIT Press, 2005.

\bibitem{blei2017variational}
David~M Blei, Alp Kucukelbir, and Jon~D McAuliffe.
\newblock Variational inference: A review for statisticians.
\newblock {\em Journal of the American statistical Association},
  112(518):859--877, 2017.

\bibitem{challis2013gaussian}
Edward Challis and David Barber.
\newblock Gaussian kullback-leibler approximate inference.
\newblock {\em The Journal of Machine Learning Research}, 14(1):2239--2286,
  2013.

\bibitem{hensman2013gaussian}
James Hensman, Nicolo Fusi, and Neil~D Lawrence.
\newblock Gaussian processes for big data.
\newblock {\em arXiv preprint arXiv:1309.6835}, 2013.

\bibitem{titsias2009variational}
Michalis Titsias.
\newblock Variational learning of inducing variables in sparse gaussian
  processes.
\newblock In {\em Artificial Intelligence and Statistics}, pages 567--574,
  2009.

\bibitem{matthews2016sparse}
Alexander G de~G Matthews, James Hensman, Richard Turner, and Zoubin
  Ghahramani.
\newblock On sparse variational methods and the kullback-leibler divergence
  between stochastic processes.
\newblock In {\em Artificial Intelligence and Statistics}, pages 231--239,
  2016.

\bibitem{bauer2016understanding}
Matthias Bauer, Mark van~der Wilk, and Carl~Edward Rasmussen.
\newblock Understanding probabilistic sparse gaussian process approximations.
\newblock In {\em Advances in neural information processing systems}, pages
  1533--1541, 2016.

\bibitem{Shi2014}
Jianing~V. Shi, Yangyang Xu, and Richard~G. Baraniuk.
\newblock {Sparse Bilinear Logistic Regression}.
\newblock pages 1--25, 2014.

\bibitem{ChristoforosChristoforou2010}
Christoforos Christoforou, {Robert Haralick}, {Paul Sajda}, and {Lucas C.
  Parra}.
\newblock {Second-Order Bilinear Discriminant Analysis}.
\newblock {\em Journal of Machine Learning Research}, 11:665--685, 2010.

\bibitem{Wood2011}
Simon~N. Wood.
\newblock {Fast stable restricted maximum likelihood and marginal likelihood
  estimation of semiparametric generalized linear models}.
\newblock {\em Journal of the Royal Statistical Society: Series B (Statistical
  Methodology)}, 73(1):3--36, jan 2011.

\bibitem{nickisch2009vglm}
Hannes Nickisch and Matthias~W. Seeger.
\newblock Convex variational bayesian inference for large scale generalized
  linear models.
\newblock In {\em Proceedings of the 26th Annual International Conference on
  Machine Learning}, ICML '09, pages 761--768. ACM, 2009.

\bibitem{hui2019semiparametric}
Francis~KC Hui, Chong You, Han~Lin Shang, and Samuel M{\"u}ller.
\newblock Semiparametric regression using variational approximations.
\newblock {\em Journal of the American Statistical Association}, pages 1--24,
  2019.

\bibitem{adam2016scalable}
Vincent Adam, James Hensman, and Maneesh Sahani.
\newblock Scalable transformed additive signal decomposition by non-conjugate
  gaussian process inference.
\newblock In {\em 2016 IEEE 26th international workshop on machine learning for
  signal processing (MLSP)}, pages 1--6. IEEE, 2016.

\bibitem{adam2017structured}
Vincent Adam.
\newblock Structured variational inference for coupled gaussian processes.
\newblock {\em arXiv preprint arXiv:1711.01131}, 2017.

\bibitem{turner2011}
R.~E. Turner and M.~Sahani.
\newblock Two problems with variational expectation maximisation for
  time-series models.
\newblock In D.~Barber, T.~Cemgil, and S.~Chiappa, editors, {\em Bayesian Time
  series models}, chapter~5, pages 109--130. Cambridge University Press, 2011.

\bibitem{Gold2007}
Joshua I. Gold and Michael N. Shadlen.
\newblock {The neural basis of decision making.}
\newblock {\em Annual review of neuroscience}, 30:535--74, jan 2007.

\bibitem{Wyart2012a}
Valentin Wyart, Vincent de~Gardelle, Jacqueline Scholl, and Christopher
  Summerfield.
\newblock {Rhythmic Fluctuations in Evidence Accumulation during Decision
  Making in the Human Brain}.
\newblock {\em Neuron}, 76(4):847--858, nov 2012.

\bibitem{DeFalguerolles2012}
Antoine de~Falguerolles.
\newblock {Generalized Multiplicative Models}.
\newblock In {\em COMPSTAT}, pages 143--175. Physica-Verlag HD, Heidelberg,
  2012.

\bibitem{Ahrens2008}
Misha~B Ahrens, J.~F. Linden, and M.~Sahani.
\newblock {Nonlinearities and Contextual Influences in Auditory Cortical
  Responses Modeled with Multilinear Spectrotemporal Methods}.
\newblock {\em Journal of Neuroscience}, 28(8):1929--1942, feb 2008.

\bibitem{Dyrholm2007}
Mads Dyrholm, Christoforos Christoforou, and Lucas~C Parra.
\newblock {Bilinear Discriminant Component Analysis}.
\newblock {\em The Journal of Machine Learning Research}, 8:1097--1111, 2007.

\bibitem{Yu2009a}
Byron~M. Yu, Jp~Cunningham, Gopal Santhanam, Si~Ryu, Krishna~V. Shenoy, and
  Maneesh Sahani.
\newblock {Gaussian-Process Factor Analysis for Low-Dimensional Single-Trial
  Analysis of Neural Population Activity}.
\newblock {\em Journal of Neurophysiology}, 102(April 2009):614--635, 2009.

\bibitem{Zhao2017}
Yuan Zhao and II~Memming Park.
\newblock {Variational Latent Gaussian Process for Recovering Single-Trial
  Dynamics from Population Spike Trains}.
\newblock {\em Neural Computation}, 29(5):1293--1316, may 2017.

\bibitem{Nadalin2019}
Jessica~K. Nadalin, Louis~Emmanuel Martinet, Ethan~B. Blackwood, Meng~Chen Lo,
  Alik~S. Widge, Sydney~S. Cash, Uri~T. Eden, and Mark~A. Kramer.
\newblock {A statistical framework to assess cross-frequency coupling while
  accounting for confounding analysis effects}.
\newblock {\em eLife}, 8, oct 2019.

\bibitem{Ehinger2018}
Benedikt~V Ehinger and Olaf Dimigen.
\newblock {Unfold: An integrated toolbox for overlap correction, non-linear
  modeling, and regression-based EEG analysis}.
\newblock {\em bioRxiv}, page 360156, dec 2018.

\end{thebibliography}
\newpage
\begin{appendices}

\section{Identifiability, constraints and offsets} \label{appendix_constraints}

One solution to the identifiability problem is to constrain all functions to take be null at some value and add offsets $\offset$. We still need to had further constraints on the offset depending on the structure of the model:
\begin{itemize}
    \item We impose $\offset=1$ for $j>1$. This avoids equivalent models by scaling all $f_{k(i1l)}$ and $c_{i1}$ by $\lambda$, and all $f_{k(ijl)}$ and $\offset$ by $1/\lambda$).
    \item If there is any fixed function in factor $j$, i.e if there is one $f_{k(ijl)}=h_{k(ijl)}$, then the offset is not needed because setting the value of this function removes the scaling equivalency. Therefore we set $\offset=0$, and remove all constraints on functions in factor $j$.
    \item we also impose $c_{i1}=0$ if $D_\c=1$. If there is no interaction terms for block $\c$, as in a standard GAM, then we need to remove the equivalence between parameters $c_{1i}$ and $c_0$.
\end{itemize}

Provided the constraints above, our model will be identifiable, unless there is a null factor, i.e. unless there is $(i,j)$ where  $\sum_l \f(x)=0$ for all $x \in \cX$. Note that it is also possible, when offset $\offset$ is not constrained, to absorbe it into one of the functions in the factor (and remove the constraint on that function). For example instead of $f_1(x)+f_2(x)+c$ with a constraint on $f_1$ and $f_2$, we can use equivalently $f_1(x)+f_2(x)$ with a constraint on $f_2$ only.

 The form of the constraint on each $f$ does not necessarily have to be that $f(x_0) =0$ for a given $x_0$. A different constraint may be used to facilitate the interpretations of the results. For example, in the experimental analysis of Figure \ref{fig:grating}, we chose a constraint that the average of the weights $w_k$ be 1. In general,  ee will use linear constraints on function evaluations $\bm p_k^T \vf_k = l_k$. By identifying an orthonormal basis $\vP_k$ for the subspace of $\mathbb R^{V}$ that is orthogonal to $\bm p_k$, we project  $\vf_k$ onto this subspace and obtain free parameters $\freepars_k = \vP_k \vf_k \sim \mathcal N(\vP_k \vmu_k, \vP_k \vK_k \vP_k^T)$, such that $\vf_k = \bm p_k l_k + \vP_k^T  \freepars_k$. If there is no constraint on a set of weights we simply have $\vP_k = \bm I$ and $l_k = 0$. In practice we will use four types of constraint:

\begin{enumerate}
   
    \item \textit{first-zero constraint}, i.e. $f_k(x_0)=0$, is the default constraint that the function must be null at some defined value $x_0$. The projection matrix $\vP_k$ is simply the identity matrix deprived of the corresponding line.

    \item \textit{mean-zero constraint}, i.e. $\sum_n f_k(n)=0$. This corresponds to $l_k = 0$ and corresponding projection matrix $P_k(m,n) = \frac{1}{\sqrt{m(m+1)}}$ if $m \geq n$, $P_k(m,m+1) = -\frac{m}{\sqrt{m(m+1)}}$ and $P_k(m,n) = 0$ if $m<n$. This constraint is useful in the GAM context, i.e. when the activations are taken as the sums of GPs $\rho(x) = \sum_k f_k(x_k)$. In \gum s, we will generally impose mean-zero constraint for all but one functions inserted in one-dimensional components. 
    
    \item \textit{mean-one constraint}, i.e. $\sum_n f_k(n) = l_k = 1$ (the projection matrix is same as for mean-zero constraint). This is equivalent to mean-zero constraint and absorbing offset $c=1$.
    
    \item \textit{sum-one constraint}, i.e. $\sum_n f_k(n)= V_k$, is an alternative to mean-one constraint.
\end{enumerate}
   

\section{Laplace approximation for \gum} 
\label{appendix_laplace}

\subsection{Conversion to GMM with Gaussian prior}

The parameters to infer are $\vtheta = \{ \vf_{1..K}, \bm c\}$, where $\vf_k= f_k(\vx_k)$ and $\vx_k$ the vector of $v_k$ unique values taken by $x_{s(k)}$ in the dataset. $f_k$ have MVN prior $\mathcal N(\vmu_k, \vK_k)$ and each offset parameter has normal prior $\mathcal N(0,\sigma^2)$, so $\vtheta$ has MVN prior with mean $(\vmu_1,..\vmu_K,\bm 0)$ and a block-diagonal covariance matrix. Instead of fully flexible functions $f_k$, we can also impose linearity, i.e. $f_k(\vx) = \bm w_k^T \vx_{s_k}$. In this case, we define an isometric MVN prior on the weights (akin to L2-regularization), i.e. $\vf_k = \bm w_k$ and $\vf_k \sim \mathcal N(\bm 0,\sigma_k^2 \bm I)$.

We can write $f_k(\vx\n) = \DM_{kn} \cdot \vtheta $ 
where $\DM_{kn}$ is an indicator vector of length $v_k$ whose value 1 indicates the position of the corresponding parameter in the parameter set. 
If function $f_k$ decomposes as the product of a fixed function $h_k$ and function to be estimated $\tilde f_k$, then parameters are $\vf_k = \tilde f_k(\vx)$ and the values of $\DM_{kn}$ are changed to $h_k(\vx\n)$. 
In the case of linear mapping, we have $\DM_k = (\vx_{s(k)})$ (i.e. the classical design matrix of a GLM). Finally a fixed function $f_k = h_k$ has no parameter.

The values taken by factor $F\ij(\vx) = \sum_l \f(\vx)$ in the dataset is $\bm F\ij = F\ij(\vX) = \DM\ij \vtheta\ij + \bm C\ij$, where $\vtheta\ij$ is the subset of $\vtheta$ that parametrises factor $F\ij$, the design matrix $\DM\ij$ for factor $F\ij$ is built by concatenating design matrices $\DM_{k(ijl)}$ for individual functions, and including also the term for dependence in offset $\offset$ if it is a free parameter. $\bm C\ij$ sums all fixed values, i.e. $\bm h_k$ for fixed functions $f_k$ in the factor, and $\offset$ if its value is fixed.

Now we see that the equation of a \gum{} (equation \ref{eq:gum}) can be replaced with a generalized multilinear model \cite{Shi2014, Ahrens2008} with $C$ blocks and Gaussian priors for the weights:  

\begin{equation}
    \vrho =  \sum_{i=1}^C \prod_{j=1}^{D_i} (\DM\ij \vtheta\ij^T + \bm C\ij)
\end{equation}

\subsection{Maximum A Posteriori weights} \label{appendix_MAP}

To identify the Maximum A Posteriori solution, we use the general solution for generalized multilinear models which is to optimise over set of parameters in one factor while keeping others factors in the block constant, pass on the next factor and iterate until convergence (\cite{Ahrens2008}). At each iteration, we optimise weights over factor $\dstar$ for each block $i$. The optimization is possible as long as the parameters $\vtheta\ij$ in the factors are not also present in the factors that are fixed, i.e. if there are no calls to the same function in the different factors of the same block (the method cannot be applied to model $\rho(\vx) = f_1(\vx)(f_1(\vx)+f_2(\vx))$. However it is perfectly possible to have different calls to the same function in different blocks, for example defining $\rho(\vx) = f_1(x_1)f_2(x_2) + f_1(x_3)f_4(x_4)$.

The generative model transforms to :

\beqarray \label{eq:GMMconstraint}
\vrho & =  & \sum_\c  (\DM_{(\c,\neg \dstar)}  \vtheta \ijstar + \vC \ijstar  )\\
   & =  & \sum_\c \DM_{(\c,\neg \dstar)} \vP\ijstar^T \freepars\ijstar + \vC\estar \text{, where} \\
   \vC\estar & = & \sum_\c (\vC \ijstar + \bm p\ijstar \Phi_{(\c,\neg \dstar)} \bm p\ijstar^T)
\eeqarray

where the new covariates are obtained by collapsing over fixed factors $\d \neq \dstar$: $\DM_{(\c,\neg \dstar)} = \prod_{\d \neq \dstar} (\DM\ij \vtheta\ij^T + \bm C\ij)$. We see that the predictor is linear with respect to the set of weights $\freepars$ corresponding to all $\freepars\ijstar$ in all blocks $\c$. We thus obtain the generative equation from a GLM with MVN prior:
\beq \label{eq:glm_prior}
\gEy = \vrho = \DMstartilde \freestarT + \vC\estar
\eeq 
where 
$\DMstartilde = \DM\estar \PstarT, \DM\estar_n = \left[ \Phi_{(1,\neg \d_1\estar)} \dots \DM_{(C,\neg \d_C\estar)} \right]$  and 
$\Pstar = \left[ 
\begin{array}{c} \vP_{(1,\d_1\estar)} 
\\ ... \\ \vP_{(C,\d_C\estar)} 
\end{array} 
\right]$

We update the set of weights $\freestar$ with a single Newton-Raphson update. Prior mean for $\freestar$ is MVN, with means $\mustar$ and covariance $\Kstar$ extracted from the $\vmu$ and $\vK$. The log-posterior over $\freestar$ can be expressed as:

\beqarray
\log p(\freestar | \vy) & = &  \log(\vy| \freestar) + \log p(\freestar) + \const \\
&  = &  \frac{1}{s} \sum_{n=1}^N (  \eta_n y\n - B(\eta_n)) - \frac{1}{2} (\Udif)^T (\Kstar)^{-1} \Udif  + \const
\eeqarray
 
 where $\eta_n$ and $s$ are respectively the canonical and dispersion parameter of the exponential family distribution for $y_n$, and $B(\eta_n)$ is such that $\frac{dB}{d\eta}=\mathbb E(y\n) = g^{-1}(\rho_n)$. In the following we assume that $g$ is the canonical link function (similar updates can be found in the general case).
 
Since the gradient of $\rho\n$ w.r.t weights $\freestar$ is $\DMstar_n \Pstar$, the gradient and Hessian of the log-posterior gives:

\beqarray   
\nabla \log p(\freestar | \vy) & = &  \frac{1}{s} \sum_{n=1}^N  \DMstartildeT_n ( y\n - g^{-1}(\rho_n) ) - (\Kstar)^{-1} \Udif  \\
\nabla \nabla \log p(\freestar | \vy) & = &  
- \frac{1}{s} \sum_{n=1}^N \DMstartildeT_n R_{nn} \DMstartilde_n  - (\Kstar)^{-1}
\eeqarray

$\bm R$ is a diagonal matrix such that $R_{nn} = (g^{-1})'\rho\n = \frac{1}{g'(g^{-1}(\rho\n))}$.

The Newton-Raphson update gives:
\beqarray
\label{eq:Newton}
\freestar_{\text{new}} & = & \freestar - (\nabla \nabla \log p(\freestar | \vy))^{-1} \nabla \log p(\freestar | \vy) \\
& = & \freestar + (\Kstar \DMstartildeT \bm R \DMstartilde + s \bm I)^{-1}(\Kstar \DMstartildeT (\vy {-} g^{-1}(\bm \rho)) - s \Udif) \\
& = & \bm H^{-1}  \bm B \text{ with} \\
\bm H & = & 
 \Kstar \DMstartildeT  \bm R \DMstartilde  +s \bm I  \\
\bm B & = &  \Kstar \DMstartildeT ( \bm R \bm {\tilde \rho}  + \vy - g^{-1}(\bm \rho)  ) +  s \mustartilde    
\eeqarray

 We have defined $\bm {\tilde \rho}=\DMstartilde \Pstar \freestar = \vrho - \vC\estar$. From equation (\ref{eq:Newton}) we obtain the new values for all weights $\vtheta_{(\c,\dstar)}$. The algorithm loops by selecting at each iteration a new set of factors $\dstar$ and then applying equations (\ref{eq:GMMconstraint},\ref{eq:glm_prior}, \ref{eq:Newton}) to update the values of $\vtheta_{(\c,\dstar)}$.\\
 
 If there are several set of unconstrained weights in the same block, convergence may take many iterations as the scaling of the weights are only constrained by the different priors. In such cases, it is convenient to re-scale these set of weights after each iteration to speed up convergence time:
 \beq
 \vtheta\ij^{\text{new}} = \frac{ (\prod_{\d'} \alpha_{\c\d')})^{\frac{1}{2D_\c^{\text{free}}} }}{\sqrt{\alpha\ij}} \vtheta\ij
 \text{ , with } 
 \alpha\ij = (\vtheta\ij - \vP\ij)^T (\Kcd)^{-1}(\vtheta\ij - \\vK\ij)^T
 \eeq

The product is taken over all free constraint dimensions in the component ($D_\c^{\text{free}}$ is the number of such dimensions).

\subsection{Posterior covariance}

In the Laplace approximation, the posterior mean is provided by the MAP weights while the posterior covariance for weights is approximated from the full Hessian of the log-posterior. Hessian for free weights in the same set (same component, same dimension) are provided by equation \ref{eq:Newton}. For free weights in different sets $\vtheta\ij$ and $\vtheta_{(\c'\d')}$, we have:

\beq
\nabla_{\freepars\ij}\nabla_{\freepars_{\c'\d'}} \log p(\freepars | \vy) = (\DM_{(\c,\neg \d)})^T \bm R \DM_{(\c',\neg \d')}
\eeq

Once we have identified the approximate posterior covariance for free parameters $\Sigmatilde = -(\nabla_{\freepars}\nabla_{\freepars} \log p(\freepars | \vy))^{-1}$, we recover the posterior for parameters $\vtheta$ which is $\vSigma = \vP^T \Sigmatilde  \vP$ (matrix $\vP$ is is block diagonal formed with all $\vP_k$ for all components and dimensions).

The Laplace approximation can be used to generate predictions for $f_k(x')$ at values of $x'$ not included in the training set (\cite{rasmussen2005}):
\beqarray
      f_k(x') & = & \mathcal N(m',(v')^2) \text{ , where} \\
      m' & = &  K_k(x', \vx_k) (\vK_k)^{-1} \vf_k\map \\
      (v')^2 & = & K_k(x',x') - K_k(x', \vx_k)^T    (\Sigma_k)^{-1}     K_k(x', \vx_k) 
\eeqarray

The Laplace approximation can also be used to approximate the log-marginal evidence $p(\vy)$ (\cite{rasmussen2005}):
\beq
\log p(\vy | \vX) \approx - \frac{1}{2} (\freepars\map-\tilde{\vmu})^T \bm {\tilde K}^{-1} (\freepars\map-\tilde{\vmu})
+ \log p(\vy |\vX, \freepars\map ) 
- \frac{1}{2} \log |\bm I +  \frac{1}{2} \bm {\tilde K} W    |
\eeq

\subsection{Hyperparameter fitting}  \label{Laplace_learning}

\subsubsection{Maximising cross-validated score}  \label{Laplace_CVLL}

Here, we first find MAP values $\freepars\map$ from a training set $(\vx, \vy)$ and compute a fitting score for $\freepars\map = arg\, max \, p(\freepars|\vx,\vy,\hp)$ on a cross-validation set $(\vX', \vy')$. We wish to find hyperparamaters $\hp$ that maximises the cross-validated score $S(\freepars; \vX', \vy')$. Here we will use the log-likelihood as the score, i.e. $S(\freepars; \vX', \vy') = \log(\vy'|\vX',\freepars)$. The gradient of the score w.r.t to hyperparameters can be computed using the chain rule: 
\beqarray
\nabla_{\hp} S(\freepars\map; \vX', \vy') & = & \bm \nabla_{\hp}  \freepars\map \cdot \nabla_{\freepars}S(\freepars\map; \vX', \vy') \\
& = & \frac{1}{s} \bm \nabla_{\hp}  \freepars\map \cdot \bm P  \sum_{k=1}^{n'} (y_{k'}-g^{-1}(a_{k'}) \nabla_{\bm U} a_{k'}
\eeqarray

From equation \ref{eq:GMMconstraint}, we can see that the gradient of $a_{k'}$ is obtaining by concatenating pseudo-design matrices $(\DM_{.k'}^{(1,\neg 1)},..,\DM_{.k'}^{(C,\neg D_C)})$ .
From the definition of the MAP weights $\freepars\map$, the Jacobian with respect to hyperparameters is:
\beqarray
    \bm \nabla_{\hp} \freepars\map  & = & - H^{-1} (\nabla_{\hp} \nabla_{\freepars} \log p(\freepars\map|\vX,\vy,\hp)) \\
    & = & - H^{-1} \bm P (\nabla_{\hp} \nabla_{\bm U} \log \mathcal N(\freepars\map; \bm 0, \vK(\hp))) \\
    & = &  -H^{-1} \bm P \vK^{-1} \nabla_{\hp} \vK \vK^{-1} \freepars\map
\eeqarray

where $H = (\nabla_{\freepars} \nabla_{\freepars} \log p(\freepars\map|\vX,\vy,\hp))$ and we have omitted the dependence of the covariance prior on the hyperparameters $\vK = \vK(\hp)$.
Instead of maximising the cross-validated score on a single cross-validation set, the score (and its gradient) can be averaged over multiple partitions of the data into training and cross-validation sets for more robust results. In many situations, the GP covariance $K$ will be close to singularity so $\vK^{-1} \nabla_{\hp} \vK \vK^{-1}$ may be subject to large numerical errors. In such case, the gradient of the score cannot be evaluated properly, so maximization of the score should use gradient-free optimization procedure such as simplex algorithms.


\subsubsection{Maximise marginal likelihood through Expectation-Maximisation}

An alternative way of fitting the hyperparameters is to maximise the marginal likelihood $p(\vy|\vX,\hp) = \int p(\vy|\vX,\freepars) p(\freepars|\hp) d\freepars$ through Expectation-Maximisation. In the E-step we use the Laplace approximation to derive an approximate posterior for the weights $p(\freepars|\vX,\hp) \approx q(\freepars) = \mathcal N(\freepars\map, \Sigmatilde)$. In the M-step we maximise the lower bound w.r.t to hyperparameters (which define covariance matrices $\vK\ij$):

\beqarray \label{eq:Mstep}
Q(\hp) & = & \int p(\freepars|\vX,\hp_{\text{old}}) \log p(\vX, \freepars | \hp) d\freepars \\
& \approx & \int \mathcal N(\freepars\map, \Sigmatilde) \log \mathcal N(\freepars; \vP, \vK) d\freepars + \text{const} \\
& \approx &  \frac{1}{2} Tr((\vP \vK \vP^T)^{-1}\Sigmatilde) + \frac{1}{2}\log \det (\vP \vK \vP^T) \\
& & + \frac{1}{2} (\freepars\map - \vP \vP)^T (\vP \vK \vP^T)^{-1}(\freepars\map - \vP \vP) + \text{const} \\
& \approx &  \frac{1}{2} \sum_k \big[ Tr((\vP_k \vK_k \vP_k^T)^{-1}\Sigmatilde_k) +
 \log \det (\vP_k \vK_k \vP_k^T)  \\
& & + (\freepars\map_k- \vP_k \muij)^T (\vP_k \vK_k \vP_k^T)^{-1}(\freepars\map_k - \vP_k \vmu_k) \big] + \text{const}
\eeqarray

The hyperparameters related to each dimension of each component $\hp_k$ can be optimised independently by maximising the related quantity in the sum of equation \ref{eq:Mstep} through gradient search. In the case of L2-regularization ($\vmu_k = \bm 0$ and $\vK_k = \lambda_k^2 \bm I$), we can get the analytical solution \cite{bishop2006pattern}:

\beq
\lambda_k^2 = \frac{||\freepars_k||^2 + Tr(\Sigmatilde_k)}{card(k)}
 \eeq

\section{Estimation error on synthetic dataset} \label{sec:measuring_error}

We measured the estimation error of GUM in different ways. 
First, we measured the mean square error of estimated predictor $\frac{1}{n}\sum_n (\hat \rho\n- \rho\n_{true})^2$. Since GUM is essentially an inference tool, where the principal interest is about inferring function $f_k$, we can also compute the error on these function. We defined the estimation error on the function $err(f)$ as the expected mean square error over function evaluation under the posterior distribution over the function, i.e. $err(f_k) = \frac{1}{n} \sum_n \int (f_k\n-f_{k,true}\n)^2 p(f_k\n|\vy) df_k\n$. The error decomposes into a bias term (the mean squared error for the posterior mean), and a variance term (the mean variance of the posterior at evaluated points):

\beqarray
err(f_k) & = &\frac{1}{n} \sum_n \int ((f-\mu_k\n)-(f_{k,true}\n-  \mu_k\n))^2 \mathcal N(f; \mu_k\n, {\sigma_k\n}^2) df \\
 & = & \frac{1}{n} \sum_n [(f\n_{k,true}-\mu_k\n)^2 + (f-\mu_k\n)^2 \\
 &&-2 (f-\mu_k\n) (f_{k,true}\n-\mu_k\n)] 
  \mathcal N(f; \mu_k\n, {\sigma_k\n}^2) df \\
 & = &  \frac{1}{n} \sum_n (f\n_{k,true}-\mu_k\n)^2 + 
 \frac{1}{n} \sum_n {\sigma_k\n}^2 
\eeqarray


\section{Experimental procedure}

 Each stimulus sequence consisted of five to ten gratings. Each grating was a high-contrast Gabor patch (colour: blue or purple; spatial frequency = 2 cycles per degree; SD of Gaussian envelope = 1 degree) presented within a circular aperture (4 degrees) against a uniform gray background. Each grating was presented during 100 ms, and the interval between gratings was fixed to 300 ms. The angles of the gratings were sampled from a von Mises distribution centered on the reference angle (45 degrees for category associated with right response, 135 degrees for category associated with left response) and with concentration coefficient $\kappa = 0.3$. Each sequence was preceded by a rectangle flashed twice during 100 ms (the interval between the flashes and between the second flash and the first grating varied between 300 and 400 ms). Participant indicated their choice with a button press after the onset of a centrally occurring dot that succeeded the backward mask and were made with a button press with the right hand. Failure to provide a response within 1000 ms after central dot onset was classified as invalid trial. Auditory feedback was provided 250 ms after participant response (at latest 1100 ms after end of stimulus sequence). It consisted of an ascending tone (400 Hz/800 Hz; 83 ms/167 ms) for correct responses; descending tone (400 Hz/ 400 Hz; 83 ms/167 ms) for incorrect responses; a low tone (400 Hz; 250 ms) for invalid trials. 
  Trials were separated by a blank interstimulus interval of 1,200-1,600 ms (truncated exponential distribution of mean 1,333 ms). Experiments consisted of 480 trials in 10 blocks of 48. It was preceded with two blocks of initiation with 36 trials each. In the first initiation block, there was only one grating in the sequence, and it was perfectly aligned with one of the reference angles. In the second initiation block, sequences of gratings were introduced, and the difficulty was gradually increased (the distribution concentration linearly decreased from  $\kappa=1.2$ to $\kappa=0.3$). 
Invalid trials (mean 6.9 per participant, std 9.4) were excluded from all regression analyses.

Visual stimuli were generated and behavioral responses recorded using Psychophysics-3 Toolbox in addition to custom scripts written for Matlab (MathWorks). 
\end{appendices}
\end{document}